\documentclass[aps,prl,twocolumn,superscriptaddress]{revtex4-1}
\usepackage{bm}
\usepackage{graphicx}
\usepackage{color}
\usepackage{hyperref}

\makeatletter

\newcommand{\Rmnum}[1]{\expandafter\@slowromancap\romannumeral #1@}
\usepackage{amsmath}

\begin{document}

\title{\textbf{Quasi-steady electron-excitonic complexes coupling in a two-dimensional semiconductor}}

\author{Shangkun Mo}
\altaffiliation{These authors contributed equally to this work.}
\affiliation{Institute for Advanced Studies, Wuhan University, Wuhan 430072, China}

\author{Hao Zhong}
\altaffiliation{These authors contributed equally to this work.}
\affiliation{Institute for Advanced Studies, Wuhan University, Wuhan 430072, China}

\author{Keming Zhao}
\affiliation{Institute for Advanced Studies, Wuhan University, Wuhan 430072, China}

\author{Yunfei Bai}
\affiliation{Beijing National Laboratory for Condensed Matter Physics and Institute of Physics, Chinese Academy of Sciences, Beijing 100190, China}
\affiliation{School of Physical Sciences, University of Chinese Academy of Sciences, Beijing 100190, China}

\author{Dingkun Qin}
\affiliation{Institute for Advanced Studies, Wuhan University, Wuhan 430072, China}

\author{Chunlong Wu}
\affiliation{Institute for Advanced Studies, Wuhan University, Wuhan 430072, China}

\author{Qiang Wan}
\affiliation{Institute for Advanced Studies, Wuhan University, Wuhan 430072, China}

\author{Renzhe Li}
\affiliation{Institute for Advanced Studies, Wuhan University, Wuhan 430072, China}

\author{Cao Peng}
\affiliation{Institute for Advanced Studies, Wuhan University, Wuhan 430072, China}

\author{Xingzhe Wang}
\affiliation{Institute for Advanced Studies, Wuhan University, Wuhan 430072, China}

\author{Enting Li}
\affiliation{Institute for Advanced Studies, Wuhan University, Wuhan 430072, China}

\author{Sheng Meng}
\affiliation{Beijing National Laboratory for Condensed Matter Physics and Institute of Physics, Chinese Academy of Sciences, Beijing 100190, China}
\affiliation{School of Physical Sciences, University of Chinese Academy of Sciences, Beijing 100190, China}
\affiliation{Songshan Lake Materials Laboratory, Dongguan, Guangdong 523808, China}

\author{Nan Xu}
\email{Corresponding author: nxu@whu.edu.cn}
\affiliation{Institute for Advanced Studies, Wuhan University, Wuhan 430072, China}
\affiliation{Wuhan Institute of Quantum Technology, Wuhan 430206, China}

\date{\today}

\begin{abstract}
Excitons and their complexes govern optical-related behaviors in semiconductors. Here, using angle-resolved photoemission spectroscopy (ARPES), we have elucidated the light-matter interaction mediated by quasi-steady excitonic complexes within a monolayer of the prototypical two-dimensional (2D) semiconductor \text{WSe}$_2$. Under continuous incident light, we have observed the generation of quasi-steady excitons and their complexes, encompassing ground and excited state excitons, trions, as well as their intricate interplay. We further show spectral evidence of electronic excitation states within the background of quasi-steady excitonic complexes, characterized by valence band (VB) effective mass renormalization, the enhanced spin-orbit coupling (SOC), the formation of an excitonic gap near the Fermi level ($E_F$) of the conduction band (CB), and intervalley excitonic band folding. Our findings not only unveil a quasi-steady excitonic complex background for the creation of diverse electronic excitations in 2D semiconductors but also offer new insights into the role of excitons in the charge density wave (CDW) formation mechanism and facilitate the advancement of correlated electronic state engineering based on the coupling between electrons and excitonic complexes in a quasi-equilibrium state.
\end{abstract}

\maketitle
Atomically thin transition metal dichalcogenides (TMDCs) have emerged as highly promising platforms for exploring novel quantum phenomena \cite{manzeli20172d, wilson2021excitons, regan2022emerging}, pivotal for advancing 2D optoelectronics and valleytronics \cite{wang2018colloquium}. Owing to their 2D nature, TMDCs exhibit weak dielectric screening, which significantly enhances the Coulomb interaction, thereby facilitating strongly bound electron-hole pairs (excitons) \cite{chernikov2014exciton}. Monolayers (MLs) of H-phase TMDCs feature valley- and spin-polarized band structures \cite{wang2018colloquium, liu2013three, riley2014direct}, enabling optically active (bright) and inactive (dark) exciton states with diverse spin-valley configurations. As exciton density increases, excitons interact with electrons, holes, and phonons, forming excitonic complexes \cite{chen2023excitonic}, which offer unique access to many-body interactions and novel phases \cite{mak2013tightly, ma2021strongly, zhang2025excitons}.

Optical spectroscopies are powerful tools for elucidating excitonic phenomena and quantifying binding energy $E_b$ \cite{wang2018colloquium, mak2013tightly}. Time-resolved angle-resolved photoemission spectroscopy with extreme ultraviolet ultrafast lasers directly probes exciton momentum and valley properties, providing insights into 2D semiconductor exciton dynamics \cite{madeo2020directly, man2021experimental, wallauer2021momentum, schmitt2022formation, mori2023spin}. Yet quasi-equilibrium excitons, their interactions with quasi-particles, and modulatory impacts on 2D electronic structures, due to the energy resolution constraints imposed by the Heisenberg uncertainty principle, remain experimentally intractable.

Here, using ARPES, we report generating and detecting quasi-steady excitonic complexes in ML \text{WSe}$_2$. We directly observe quasi-steady bright/dark excitons, whose properties ($E_b$, root mean square radius $R_{\mathrm{RMS}}$, valley location) align with ultrafast ARPES measurements \cite{madeo2020directly, man2021experimental}, validating our approach. By modulating exciton density via surface electron doping, we access excitonic/trionic excited states and resolve their interactions in momentum space. We reveal that electron–excitonic complexes coupling in ML \text{WSe}$_2$ not only profoundly modulates the electronic structure but also induces the emergence of diverse quasi-particles. These findings broaden understanding of excitonic effects in 2D materials, highlight light–matter interactions as a key tool for investigating correlated quantum phenomena, and advance the field of 2D exciton science.

\begin{figure}[t!]
\centering
\includegraphics[width=\columnwidth]{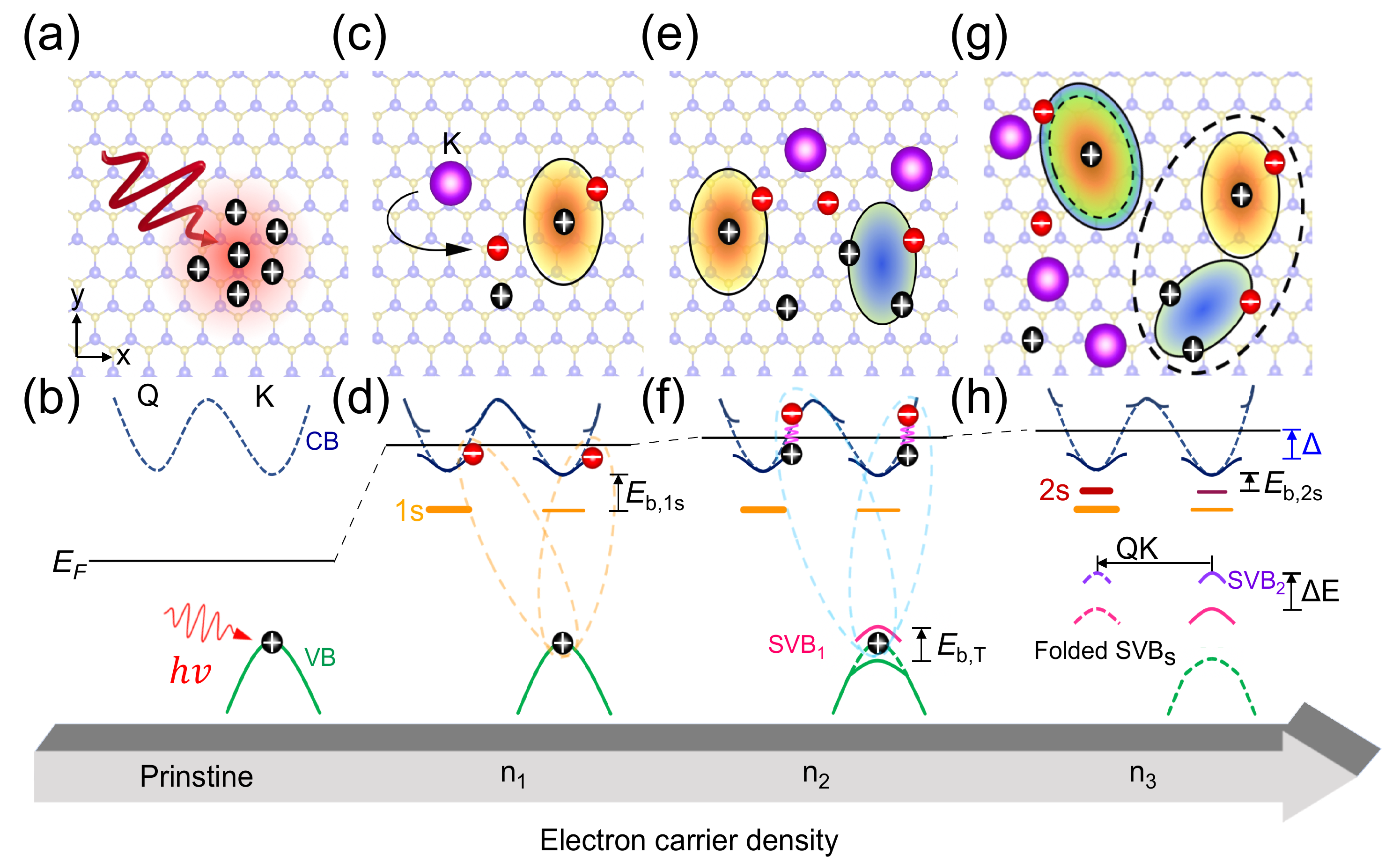}
\caption{(a-b) Photo-generated holes in ARPES measurements on pristine \text{WSe}$_2$ ML in real and momentum space, respectively. (c) Excitons are formed by photo-generated holes and doped electrons. (d) The orange lines represent the ARPES signatures of bright and dark excitons, respectively. (e) Trions are formed by photo-generated holes and electron-hole excitations near $E_F$. (f) The \text{SVB}$_1$ represents the ARPES signature of trions, and the VB edge undergoes renormalization. (g) Excited 2s-states of exciton emerges, accompanied by exciton-trion interactions. (h) The red lines and \text{SVB}$_2$ represent the ARPES signatures of exciton 2s-states and exciton-trion interactions, respectively. Folded SVBs emerge at the Q-valley.}
\label{fig:figure1}
\end{figure}

High quality \text{WSe}$_2$ MLs were synthesized by molecular beam epitaxy, on the bilayer graphene/\text{SiC} substrate \cite{zhang2016electronic}. The potassium (K) atoms were in situ deposited on clean surfaces of \text{WSe}$_2$ MLs at T=6 K. Data were acquired in a vacuum better than $5\times10^{-11}$ mbar with energy/angular resolution of 40 meV and 0.1°. $E_F$ does not vary with doping. It was determined by measuring an annealed tantalum reference in electrical contact with the samples. ARPES measurements were conducted with light from a helium lamp source with $h\nu$ = 21.2 eV.

\begin{figure*}[t!]
\centering
\includegraphics[width=2\columnwidth]{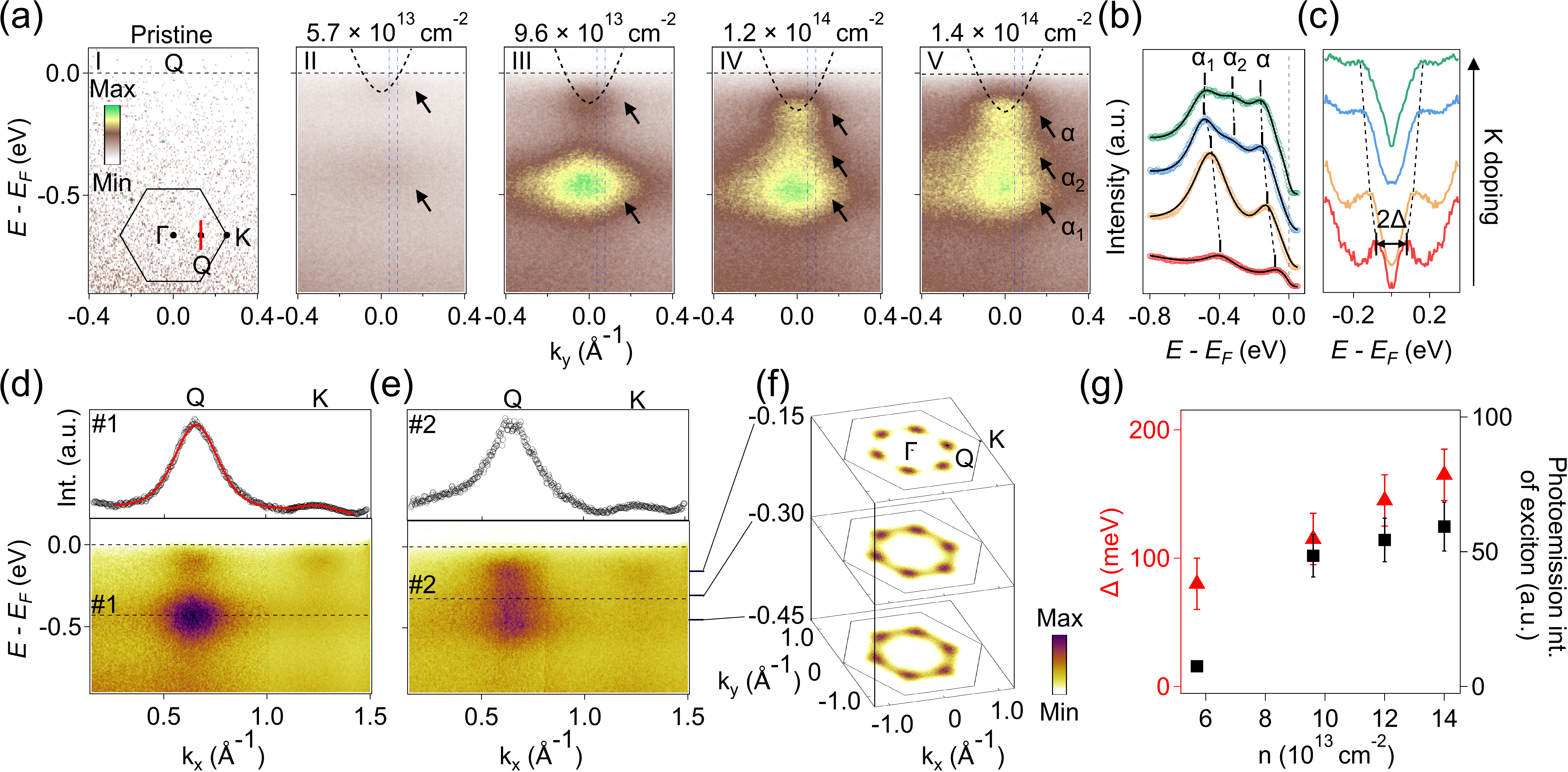}
\caption{(a) Doping-dependent ARPES intensity plots along the red line direction, with the bare CB appended. The CB, 1s-state and 2s-state of dark excitons are denoted as $\alpha$, $\alpha_1$ and $\alpha_2$, respectively. (b) Doping-dependent EDCs at the Q-point and fitting analysis. (c) Doping-dependent symmetrized EDCs at $k_F$ (blue dashed lines in (a)). The energy gap $\Delta$ is marked by the arrow. (d) ARPES intensity plot for $n = 9.6 \times 10^{13} \, \text{cm}^{-2}$ along the $\Gamma$-K direction, with \text{MDC}$\#1$ of the 1s-states of excitons and fitting analysis. (e) Plot similar to (d) but for $n = 1.4 \times 10^{14} \, \text{cm}^{-2}$, with \text{MDC}$\#2$ of the 2s-states of excitons. (f) Symmetrized constant energy maps. (g) Doping-dependent gap value $\Delta$ and the photoemission intensity of excitons, respectively.}
\label{fig:figure2}
\end{figure*}

For clarity, Fig. 1 illustrates the mechanism of quasi-steady excitonic complexes probed by ARPES and our key findings. In a pristine 2D semiconductor, photoelectron emission from the surface generates holes in the VB (Figs. 1(a–b)), which significantly influence the electronic structure, especially at low temperatures \cite{wan2025sensitive, katoch2018giant, ma2022}. Surface K deposition tunes the chemical potential by introducing electron carriers (Fig. 1(c)), shifting the CB below $E_F$ (Fig. 1(d)). Doped electrons and photo-generated holes bind via Coulomb interaction to form excitons, yielding ARPES sidebands below the CB edges (Fig. 1(d)) \cite{rustagi2018photoemission, mo2025observation}. The energy interval between the CB edge and sideband corresponds to the ground-state exciton binding energy $E_{b,\mathrm{1s}}$. Notably, the pairing between photo-generated holes and electrons in the K-/Q-valleys produces bright/dark excitons, respectively, with the latter exhibiting a longer lifetime \cite{chen2023excitonic}. During photoemission, photo-generated holes interact with Fermi sea excitations near $E_F$ to form trions (Fig. 1(e)) \cite{katoch2018giant}, producing an additional photoemission signal from side valence band 1 (\text{SVB}$_1$) above the VB edge. The energy interval between VB and \text{SVB}$_1$ corresponds to the trion binding energy $E_{b,\mathrm{T}}$ (Fig. 1(f)). Further increasing the exciton density populates the excited 2s-states of excitons, manifested as extra sidebands below the CB with reduced binding energy $E_{b,\mathrm{2s}}$ (Figs. 1(g–h)). Trion–exciton interactions (Fig. 1(g)) give rise to trion side valence band 2 (\text{SVB}$_2$), featuring an energy gain $\Delta E = E_{b,\mathrm{1s}}$ (Fig. 1(h)). We observe that the electronic structure of the \text{WSe}$_2$ ML is strongly modulated by excitons, with a gap $\Delta$ opening in the CB near $E_F$, consistent with prior studies on quasi-steady excitons in 3D semiconductors \cite{mo2025observation}, and the effective mass of the VB being altered (Fig. 1(f)). Owing to the multi-valley nature of \text{WSe}$_2$, interactions between excitonic complexes and electronic quasiparticles involve intervalley momentum transfer $\vec{\mathrm{QK}}$, which gives rise to the emergence of folded SVBs in Q-valleys (Fig. 1(h)).

Subsequently, we report ARPES results that conclusively validate the mechanism depicted in Fig. 1. Figure 2(a) illustrates the K-induced evolution of the band structure near the Q-valley in \text{WSe}$_2$ ML. In the pristine state, the energy window within the band gap shows no spectral features (Fig. 2(a-I)). Upon K deposition, the CB $\alpha$ edge shifts below $E_F$, corresponding to an electron density of $n = 5.7 \times 10^{13} \, \text{cm}^{-2}$ calculated using Luttinger theory (see SM \cite{SI} Sec. S2). Besides the CB $\alpha$ edge near $E_F$ in Fig. 2(a-II), an additional energy spectral feature $\alpha_1$ emerges, which we assign to the momentum-forbidden dark exciton sideband shown in Fig. 1(d). The spectral weight of $\alpha_1$ increases with higher electron density ($n = 9.6 \times 10^{13} \, \text{cm}^{-2}$, Fig. 2(a-III)), demonstrating controllable exciton density via surface doping. Along the $\Gamma$-K direction in the K-valley, we detect the bright exciton and its sideband, which is weaker by approximately an order of magnitude than the Q-valley dark exciton (\text{MDC}$\#1$ in Fig. 2(d)), consistent with the dark exciton’s longer lifetime due to reduced decay rates \cite{wang2018colloquium, chen2023excitonic, chand2023interaction}. Analyzing integrated energy distribution curves (EDCs) and momentum distribution curves (MDCs) (Figs. 2(b) and 2(d)), we determine exciton parameters\cite{rustagi2018photoemission}: both dark and bright excitons exhibit a binding energy $E_{b,\mathrm{1s}}$ = 320 meV, with RMS radii $R_{\mathrm{RMS, x}}$ of 1.67 nm and 1.62 nm, respectively. These values align with theoretical predictions \cite{wu2019exciton}, ultrafast ARPES measurements \cite{madeo2020directly, man2021experimental}, and optical spectroscopy results \cite{he2014tightly}, validating ARPES for probing quasi-equilibrium excitons. Compared to ML \text{WSe}$_2$/\text{hBN} heterostructures ($\mathrm{\sim400\ meV}$), the stronger screening from the bilayer graphene substrate slightly decreases $E_{b}$.

Upon increasing the carrier density to $n = 1.4 \times 10^{14} \text{cm}^{-2}$, we observe additional spectral weight between the CB $\alpha$ and the exciton sideband $\alpha_1$ (Figs. 2(a–IV) and (a-V)). This feature is attributed to the $\alpha_2$ sideband from the excited 2s exciton state, as shown in Fig. 1(g–h). Integrated EDC analysis in Fig. 2(b) gives an experimental binding energy $E_{b,\mathrm{2s}}$ of 160 meV for the 2s state, with $E_{b,\mathrm{1s}} / E_{b,\mathrm{2s}} = 2$ agreeing well with the 1.85 value for \text{WSe}$_2$/\text{SiO}$_2$ heterostructure \cite{he2014tightly}. As detailed in SM \cite{SI} Sec. S5, the exciton binding energy remains stable with electron doping, excluding a plasmonic origin since plasmon energy typically increases with carrier density \cite{riley2018crossover, ulstrup2024observation}. At the K-valley, the 2s exciton state is also detected, exhibiting comparable binding energy but reduced spectral weight, as seen in integrated \text{MDC}$\#2$ along $\Gamma$–K (Fig. 2(e)). Momentum maps of dark exciton ground and excited states reveal six-fold symmetry around $\Gamma$, resolvable from the $\alpha_1$ and $\alpha_2$ sidebands in Fig. 2(f). The six-fold symmetry of CB and dark excitons (SM \cite{SI} Sec. S6) explicitly excludes mid-gap impurity states \cite{zhang2014observation} or an impurity band \cite{Strocov2018}, as potassium atoms exhibit random disordered distribution on the sample surface.

Beyond excitonic sidebands, we observe gradual suppression of CB $\alpha$ spectral weight at $E_F$ (Fig. 2(a)), indicating energy gap opening. Figure 2(c) shows the symmetrized EDC at the Fermi momentum $k_F$ of the $\alpha$ band, with gap value $\Delta$ determined via double-peak separation. The gap evolution in Fig. 2(g) correlates with exciton photoemission intensity from EDCs in Fig. 2(b), showing exciton-mediated single-particle band renormalization. This matches the excitonic gap phase in three-dimensional semiconductor \text{SnSe}$_2$ \cite{mo2025observation}. The gap arises from photo-generated hole-mediated electron-electron interactions, driving correlated electronic state emergence.

We now focus on the VB evolution induced by excitonic complexes. Figures 3(a) and 3(b) present the doping-dependent band dispersions and their second derivatives along the $\Gamma$-K direction. Surface K-doping of the pristine \text{WSe}$_2$ ML causes rapid downward shifts of VBs $\beta$ and $\gamma$ (Figs. 3(a-I) and (a-II)) since $E_F$ resides within the semiconductor band gap. Once the CB $\alpha$ edge reaches $E_F$ (Fig. 3(a-II)), the VBs shift more slowly with increasing K-deposition time, a trend similar to CB evolution shown in Fig. 2a. At a carrier density of $n = 9.6 \times 10^{13} \text{cm}^{-2}$, an additional SVB $\beta_1$ emerges above the original $\beta$ band edge (Fig. 3(a-III)), clearly resolved in the integrated EDC plot of Fig. 3(c). The appearance of SVB $\beta_1$ stems from interactions between photo-generated holes from the $\beta$ band and $E_F$-nearby electron-hole excitations, driving trion formation as shown in Figs. 1(e) and 1(f). The trion binding energy, derived from the effective spin-orbit splitting difference between $\beta$-$\gamma$ and $\beta_1$-$\gamma$, is $\Delta E$ = $E_{b,\mathrm{T}}$ = $\Delta_\mathrm{SOC}(\beta_1-\gamma) - \Delta_\mathrm{SOC}(\beta-\gamma) =$ 210\ meV, consistent with prior results \cite{katoch2018giant}. Additionally, effective mass renormalization at the $\beta$ band top reflects strong excitonic modulation of the electronic states in the \text{WSe}$_2$ ML.

At higher doping levels, spectral weight of the original VB $\mathrm{\beta}$ transfers to the SVB $\mathrm{\beta_1}$ as increased photo-generated holes interact with $E_F$-nearby excitations to form trions. The spin-orbit splitting $\Delta_\mathrm{SOC}$({$\beta$-$\gamma$}) remains nearly unchanged across the electron carrier density range from $n = 7.0 \times 10^{13} \, \text{cm}^{-2}$ to $n = 1.2 \times 10^{14} \, \text{cm}^{-2}$ (see SM \cite{SI} Sec. S9). Figure 3(e) shows the extracted carrier-density-dependent $E_{b,\mathrm{T}}$. The agreement between $E_{b,\mathrm{T}}$ and 2$\Delta$ within error margin demonstrates light-matter interaction mediated by excitonic complexes. As illustrated in Fig. 1(f), the energy gap opening at $E_F$ sets a 2$\Delta$ threshold for Fermi-sea excitations.

\begin{figure*}[t!]
\centering
\includegraphics[width=2\columnwidth]{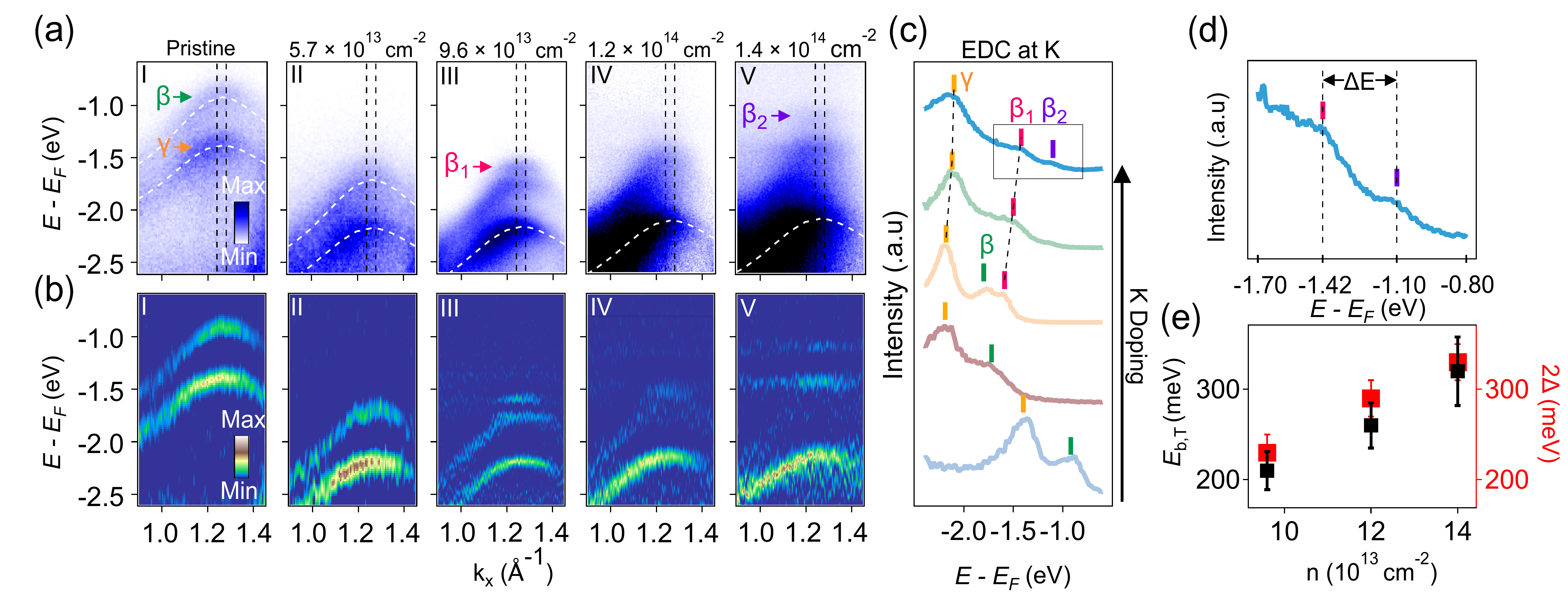}
\caption{(a) Doping-dependent VBs around the K-valley along the $\Gamma$-K direction, with the bare VBs appended. The VBs $\beta$, $\gamma$, and SVBs $\beta_1$, $\beta_2$ are marked by arrows, respectively. (b) Corresponding second-derivatives of (a). (c) Doping-dependent EDCs at the K point, with the energy positions of VBs and SVBs marked, respectively. (d) Enlarged view of the box in (c). The energy interval $\Delta E $ between SVBs $\beta_1$ and $\beta_2$ is marked by the arrow. (e) $E_{b,\mathrm{T}}$ and 2$\Delta$ (extracted from Fig. 2(g)), are plotted as a function of the electron carrier density, respectively. All ARPES plots are background-subtracted for clarity, with raw data available in SM \cite{SI} Sec. S10.}
\label{Fig3_C}
\end{figure*}

At a carrier density of $n = 1.4 \times 10^{14} \, \text{cm}^{-2}$, an additional side valence band SVB $\beta_2$ emerges above SVB $\beta_1$ (Fig. 3(a-V)). The energy difference between $\beta_2$ and $\beta_1$ closely matches the exciton 1s-state binding energy, with $\Delta E = E_{b,\mathrm{1s}}$= 320 meV (Fig. 3(d)). We assign the appearance of $\beta_2$ band to interactions between trions and exciton 1s-states, as shown in Figs. 1(g–h). Here, SVB $\beta_1$ gains the extra energy $ E_{b,\mathrm{1s}}$ from excitons, forming replica band $\beta_2$. 

\begin{figure}[t!]
\centering
\includegraphics[width=1\columnwidth]{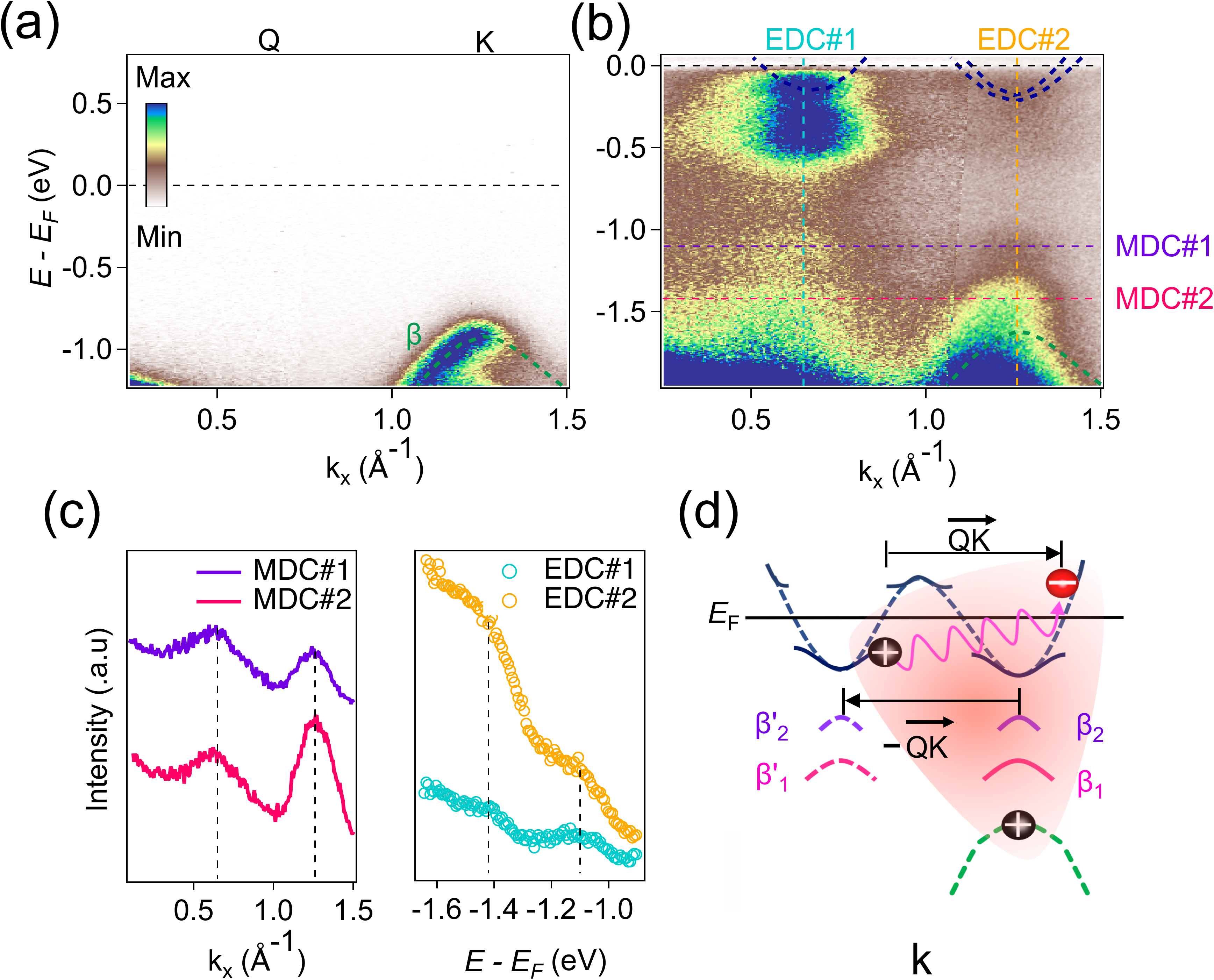}
\caption{(a) ARPES intensity plot along the $\Gamma-K$ direction for pristine sample, with the bare band $\beta$ appended. (b) Plot similar to (a), but for $n = 1.4 \times 10^{14} \, \text{cm}^{-2}$, with background subtracted (for details, see SM \cite{SI} Sec. S11). (c) MDCs at the energies of SVBs $\beta_1$ and $\beta_2$, as indicated by the horizontal lines in (b). EDCs at the Q-valley and K-valley, respectively. Dashed lines mark the peak positions. (d) Illustration of intervalley electron-hole coupling with total momentum transfer $\vec{\mathrm{QK}}$, mediated by dressed photo-generated holes at the K-valley VB $\beta$, inducing the folding of SVBs $\beta_1$ and $\beta_2$ from the K- to the Q-valley.}	
\label{Fig4_D}
\end{figure}

In addition to the gap opening in CB $\alpha$ and VB $\beta$ renormalization, we observe intervalley band folding of SVBs $\beta_1$ and $\beta_2$ from the K to Q valley. Figures 4(a) and 4(b) depict the band structures of the pristine sample and $n = 1.4 \times 10^{14} \text{cm}^{-2}$ (measured from a higher-quality sample across extended momentum/energy ranges). In Fig. 4(b), hole-like bands $\beta_1'$ and $\beta_2'$ appear near the Q-valley, features absent in the pristine sample. MDC analysis confirms their Q-valley location, with energy positions analyzed from EDCs matching those of SVBs $\beta_1$ and $\beta_2$, validating their origin from K-to-Q valley band folding (Fig. 4(c)).

The intervalley band folding of SVBs in \text{WSe}$_2$ ML originates from the material’s multivalley characteristics and electron–excitonic coupling. One proposed mechanism involves electron–hole excitations transferring from Q- to K-valley with a total momentum $\vec{\mathrm{QK}}$, as shown in Fig. 4(d). Trion formation via these finite-momentum excitations requires momentum conservation, driving SVB folding from K- to Q-valley. Electronic reconstructions such as CDW with wave vector $\vec{\mathrm{QK}}$ could also yield folded SVBs. In excitonic insulators, exciton condensation typically induces gap opening near $E_F$ and CDW formation \cite{cercellier2007evidence, gao2024observation, monney2009spontaneous}. Our observation of excitonic complexes and a gap near $E_F$ in doped 2D semiconductors suggests that exciton-driven CDW-like reconstructions may explain the folded SVBs.

In summary, we demonstrate that conventional ARPES directly detects multiple excitonic complexes in ML \text{WSe}$_2$, including ground/excited state bright/dark excitons, trions, and their interactions. Table 1 summarizes experimentally determined key parameters ($E_b$ and $R_{\mathrm{RMS}}$) of these excitonic complexes (see SM \cite{SI} Sec. S14). Moreover, the large energy scales of exciton/trion binding energies definitively exclude bosonic modes such as phonon coupling, which is incompatible with the observed energy scale \cite{mishra2018exciton}. Compared to our previous work \cite{mo2025observation} and ultrafast excitonic studies, this research not only reports the first observation of quasi-steady-state excitons in 2D semiconductors but also quantifies their modulation effects on electronic structures and the interactions between excitonic complexes and quasiparticles. Specifically, we reveal that strong electron–exciton coupling induces changes in the electronic structure, manifested as CB gap opening, VB mass renormalization, enhanced SOC, and K-to-Q SVB band folding. Collectively, excitonic complex formation, gap phase, and SVB folding suggest intervalley order mediated by excitons, consistent with CDW-like electronic reconstructions. Our findings establish a methodology to engineer quasi-steady excitonic backgrounds in 2D semiconductors, clarify the underlying interaction mechanisms, and enable tunable control of electronic structures and CDW-like states via electron–exciton complex coupling.

\textit{Data availability}—All data are processed by Igor Pro 8.0 software. All data generated during the current study are available from the corresponding author upon request.

\bibliography{wse2}

\begin{thebibliography}{34}%
\makeatletter
\providecommand \@ifxundefined [1]{%
 \@ifx{#1\undefined}
}%
\providecommand \@ifnum [1]{%
 \ifnum #1\expandafter \@firstoftwo
 \else \expandafter \@secondoftwo
 \fi
}%
\providecommand \@ifx [1]{%
 \ifx #1\expandafter \@firstoftwo
 \else \expandafter \@secondoftwo
 \fi
}%
\providecommand \natexlab [1]{#1}%
\providecommand \enquote  [1]{``#1''}%
\providecommand \bibnamefont  [1]{#1}%
\providecommand \bibfnamefont [1]{#1}%
\providecommand \citenamefont [1]{#1}%
\providecommand \href@noop [0]{\@secondoftwo}%
\providecommand \href [0]{\begingroup \@sanitize@url \@href}%
\providecommand \@href[1]{\@@startlink{#1}\@@href}%
\providecommand \@@href[1]{\endgroup#1\@@endlink}%
\providecommand \@sanitize@url [0]{\catcode `\\12\catcode `\$12\catcode
  `\&12\catcode `\#12\catcode `\^12\catcode `\_12\catcode `\%12\relax}%
\providecommand \@@startlink[1]{}%
\providecommand \@@endlink[0]{}%
\providecommand \url  [0]{\begingroup\@sanitize@url \@url }%
\providecommand \@url [1]{\endgroup\@href {#1}{\urlprefix }}%
\providecommand \urlprefix  [0]{URL }%
\providecommand \Eprint [0]{\href }%
\providecommand \doibase [0]{http://dx.doi.org/}%
\providecommand \selectlanguage [0]{\@gobble}%
\providecommand \bibinfo  [0]{\@secondoftwo}%
\providecommand \bibfield  [0]{\@secondoftwo}%
\providecommand \translation [1]{[#1]}%
\providecommand \BibitemOpen [0]{}%
\providecommand \bibitemStop [0]{}%
\providecommand \bibitemNoStop [0]{.\EOS\space}%
\providecommand \EOS [0]{\spacefactor3000\relax}%
\providecommand \BibitemShut  [1]{\csname bibitem#1\endcsname}%
\let\auto@bib@innerbib\@empty
\bibitem [{\citenamefont {Manzeli}\ \emph {et~al.}(2017)\citenamefont
  {Manzeli}, \citenamefont {Ovchinnikov}, \citenamefont {Pasquier},
  \citenamefont {Yazyev},\ and\ \citenamefont {Kis}}]{manzeli20172d}%
  \BibitemOpen
  \bibfield  {author} {\bibinfo {author} {\bibfnamefont {S.}~\bibnamefont
  {Manzeli}}, \bibinfo {author} {\bibfnamefont {D.}~\bibnamefont
  {Ovchinnikov}}, \bibinfo {author} {\bibfnamefont {D.}~\bibnamefont
  {Pasquier}}, \bibinfo {author} {\bibfnamefont {O.V.}\ \bibnamefont {Yazyev}},
  \ and\ \bibinfo {author} {\bibfnamefont {A.}~\bibnamefont {Kis}},\ }\bibfield
   {title} {\enquote {\bibinfo {title} {2{D} transition metal
  dichalcogenides},}\ }\href@noop {} {\bibfield  {journal} {\bibinfo  {journal}
  {Nat. Rev. Mater.}\ }\textbf {\bibinfo {volume} {2}},\ \bibinfo {pages}
  {1--15} (\bibinfo {year} {2017})}\BibitemShut {NoStop}%
\bibitem [{\citenamefont {Wilson}\ \emph {et~al.}(2021)\citenamefont {Wilson},
  \citenamefont {Yao}, \citenamefont {Shan},\ and\ \citenamefont
  {Xu}}]{wilson2021excitons}%
  \BibitemOpen
  \bibfield  {author} {\bibinfo {author} {\bibfnamefont {N.P.}\ \bibnamefont
  {Wilson}}, \bibinfo {author} {\bibfnamefont {W.}~\bibnamefont {Yao}},
  \bibinfo {author} {\bibfnamefont {J.}~\bibnamefont {Shan}}, \ and\ \bibinfo
  {author} {\bibfnamefont {X.}~\bibnamefont {Xu}},\ }\bibfield  {title}
  {\enquote {\bibinfo {title} {Excitons and emergent quantum phenomena in
  stacked 2d semiconductors},}\ }\href@noop {} {\bibfield  {journal} {\bibinfo
  {journal} {Nature}\ }\textbf {\bibinfo {volume} {599}},\ \bibinfo {pages}
  {383--392} (\bibinfo {year} {2021})}\BibitemShut {NoStop}%
\bibitem [{\citenamefont {Regan}\ \emph {et~al.}(2022)\citenamefont {Regan},
  \citenamefont {Wang}, \citenamefont {Paik}, \citenamefont {Zeng},
  \citenamefont {Zhang}, \citenamefont {Zhu}, \citenamefont {MacDonald},
  \citenamefont {Deng},\ and\ \citenamefont {Wang}}]{regan2022emerging}%
  \BibitemOpen
  \bibfield  {author} {\bibinfo {author} {\bibfnamefont {E.C.}\ \bibnamefont
  {Regan}}, \bibinfo {author} {\bibfnamefont {D.}~\bibnamefont {Wang}},
  \bibinfo {author} {\bibfnamefont {E.Y.}\ \bibnamefont {Paik}}, \bibinfo
  {author} {\bibfnamefont {Y.}~\bibnamefont {Zeng}}, \bibinfo {author}
  {\bibfnamefont {L.}~\bibnamefont {Zhang}}, \bibinfo {author} {\bibfnamefont
  {J.}~\bibnamefont {Zhu}}, \bibinfo {author} {\bibfnamefont {A.H.}\
  \bibnamefont {MacDonald}}, \bibinfo {author} {\bibfnamefont {H.}~\bibnamefont
  {Deng}}, \ and\ \bibinfo {author} {\bibfnamefont {F.}~\bibnamefont {Wang}},\
  }\bibfield  {title} {\enquote {\bibinfo {title} {Emerging exciton physics in
  transition metal dichalcogenide heterobilayers},}\ }\href@noop {} {\bibfield
  {journal} {\bibinfo  {journal} {Nat. Rev. Mater.}\ }\textbf {\bibinfo
  {volume} {7}},\ \bibinfo {pages} {778--795} (\bibinfo {year}
  {2022})}\BibitemShut {NoStop}%
\bibitem [{\citenamefont {Wang}\ \emph {et~al.}(2018)\citenamefont {Wang},
  \citenamefont {Chernikov}, \citenamefont {Glazov}, \citenamefont {Heinz},
  \citenamefont {Marie}, \citenamefont {Amand},\ and\ \citenamefont
  {Urbaszek}}]{wang2018colloquium}%
  \BibitemOpen
  \bibfield  {author} {\bibinfo {author} {\bibfnamefont {G.}~\bibnamefont
  {Wang}}, \bibinfo {author} {\bibfnamefont {A.}~\bibnamefont {Chernikov}},
  \bibinfo {author} {\bibfnamefont {M.M.}\ \bibnamefont {Glazov}}, \bibinfo
  {author} {\bibfnamefont {T.F.}\ \bibnamefont {Heinz}}, \bibinfo {author}
  {\bibfnamefont {X.}~\bibnamefont {Marie}}, \bibinfo {author} {\bibfnamefont
  {T.}~\bibnamefont {Amand}}, \ and\ \bibinfo {author} {\bibfnamefont
  {B.}~\bibnamefont {Urbaszek}},\ }\bibfield  {title} {\enquote {\bibinfo
  {title} {Colloquium: Excitons in atomically thin transition metal
  dichalcogenides},}\ }\href@noop {} {\bibfield  {journal} {\bibinfo  {journal}
  {Rev. Mod. Phys.}\ }\textbf {\bibinfo {volume} {90}},\ \bibinfo {pages}
  {021001} (\bibinfo {year} {2018})}\BibitemShut {NoStop}%
\bibitem [{\citenamefont {Chernikov}\ \emph {et~al.}(2014)\citenamefont
  {Chernikov}, \citenamefont {Berkelbach}, \citenamefont {Hill}, \citenamefont
  {Rigosi}, \citenamefont {Li}, \citenamefont {Aslan}, \citenamefont
  {Reichman}, \citenamefont {Hybertsen},\ and\ \citenamefont
  {Heinz}}]{chernikov2014exciton}%
  \BibitemOpen
  \bibfield  {author} {\bibinfo {author} {\bibfnamefont {A.}~\bibnamefont
  {Chernikov}}, \bibinfo {author} {\bibfnamefont {T.C.}\ \bibnamefont
  {Berkelbach}}, \bibinfo {author} {\bibfnamefont {H.M.}\ \bibnamefont {Hill}},
  \bibinfo {author} {\bibfnamefont {A.}~\bibnamefont {Rigosi}}, \bibinfo
  {author} {\bibfnamefont {Y.}~\bibnamefont {Li}}, \bibinfo {author}
  {\bibfnamefont {B.}~\bibnamefont {Aslan}}, \bibinfo {author} {\bibfnamefont
  {D.R.}\ \bibnamefont {Reichman}}, \bibinfo {author} {\bibfnamefont {M.S.}\
  \bibnamefont {Hybertsen}}, \ and\ \bibinfo {author} {\bibfnamefont {T.F.}\
  \bibnamefont {Heinz}},\ }\bibfield  {title} {\enquote {\bibinfo {title}
  {Exciton binding energy and nonhydrogenic rydberg series in monolayer
  {WS}$_2$},}\ }\href@noop {} {\bibfield  {journal} {\bibinfo  {journal} {Phys.
  Rev. Lett.}\ }\textbf {\bibinfo {volume} {113}},\ \bibinfo {pages} {076802}
  (\bibinfo {year} {2014})}\BibitemShut {NoStop}%
\bibitem [{\citenamefont {Liu}\ \emph {et~al.}(2013)\citenamefont {Liu},
  \citenamefont {Shan}, \citenamefont {Yao}, \citenamefont {Yao},\ and\
  \citenamefont {Xiao}}]{liu2013three}%
  \BibitemOpen
  \bibfield  {author} {\bibinfo {author} {\bibfnamefont {G.B.}\ \bibnamefont
  {Liu}}, \bibinfo {author} {\bibfnamefont {W.Y.}\ \bibnamefont {Shan}},
  \bibinfo {author} {\bibfnamefont {Y.}~\bibnamefont {Yao}}, \bibinfo {author}
  {\bibfnamefont {W.}~\bibnamefont {Yao}}, \ and\ \bibinfo {author}
  {\bibfnamefont {D.}~\bibnamefont {Xiao}},\ }\bibfield  {title} {\enquote
  {\bibinfo {title} {Three-band tight-binding model for monolayers of group-vib
  transition metal dichalcogenides},}\ }\href@noop {} {\bibfield  {journal}
  {\bibinfo  {journal} {Phys. Rev. B}\ }\textbf {\bibinfo {volume} {88}},\
  \bibinfo {pages} {085433} (\bibinfo {year} {2013})}\BibitemShut {NoStop}%
\bibitem [{\citenamefont {Riley}\ \emph {et~al.}(2014)\citenamefont {Riley},
  \citenamefont {Mazzola}, \citenamefont {Dendzik}, \citenamefont {Michiardi},
  \citenamefont {Takayama}, \citenamefont {Bawden}, \citenamefont {Granerød},
  \citenamefont {Leandersson}, \citenamefont {Balasubramanian}, \citenamefont
  {Hoesch} \emph {et~al.}}]{riley2014direct}%
  \BibitemOpen
  \bibfield  {author} {\bibinfo {author} {\bibfnamefont {J.M.}\ \bibnamefont
  {Riley}}, \bibinfo {author} {\bibfnamefont {F.}~\bibnamefont {Mazzola}},
  \bibinfo {author} {\bibfnamefont {M.}~\bibnamefont {Dendzik}}, \bibinfo
  {author} {\bibfnamefont {M.}~\bibnamefont {Michiardi}}, \bibinfo {author}
  {\bibfnamefont {T.}~\bibnamefont {Takayama}}, \bibinfo {author}
  {\bibfnamefont {L.}~\bibnamefont {Bawden}}, \bibinfo {author} {\bibfnamefont
  {C.}~\bibnamefont {Granerød}}, \bibinfo {author} {\bibfnamefont
  {M.}~\bibnamefont {Leandersson}}, \bibinfo {author} {\bibfnamefont
  {T.}~\bibnamefont {Balasubramanian}}, \bibinfo {author} {\bibfnamefont
  {M.}~\bibnamefont {Hoesch}},  \emph {et~al.},\ }\bibfield  {title} {\enquote
  {\bibinfo {title} {Direct observation of spin-polarized bulk bands in an
  inversion-symmetric semiconductor},}\ }\href@noop {} {\bibfield  {journal}
  {\bibinfo  {journal} {Nat. Phys.}\ }\textbf {\bibinfo {volume} {10}},\
  \bibinfo {pages} {835--839} (\bibinfo {year} {2014})}\BibitemShut {NoStop}%
\bibitem [{\citenamefont {Chen}\ \emph {et~al.}(2023)\citenamefont {Chen},
  \citenamefont {Lian}, \citenamefont {Meng}, \citenamefont {Ma},\ and\
  \citenamefont {Shi}}]{chen2023excitonic}%
  \BibitemOpen
  \bibfield  {author} {\bibinfo {author} {\bibfnamefont {X.}~\bibnamefont
  {Chen}}, \bibinfo {author} {\bibfnamefont {Z.}~\bibnamefont {Lian}}, \bibinfo
  {author} {\bibfnamefont {Y.}~\bibnamefont {Meng}}, \bibinfo {author}
  {\bibfnamefont {L.}~\bibnamefont {Ma}}, \ and\ \bibinfo {author}
  {\bibfnamefont {S.F.}\ \bibnamefont {Shi}},\ }\bibfield  {title} {\enquote
  {\bibinfo {title} {Excitonic complexes in two-dimensional transition metal
  dichalcogenides},}\ }\href@noop {} {\bibfield  {journal} {\bibinfo  {journal}
  {Nat. Commun.}\ }\textbf {\bibinfo {volume} {14}},\ \bibinfo {pages} {8233}
  (\bibinfo {year} {2023})}\BibitemShut {NoStop}%
\bibitem [{\citenamefont {Mak}\ \emph {et~al.}(2013)\citenamefont {Mak},
  \citenamefont {He}, \citenamefont {Lee}, \citenamefont {Lee}, \citenamefont
  {Hone}, \citenamefont {Heinz},\ and\ \citenamefont {Shan}}]{mak2013tightly}%
  \BibitemOpen
  \bibfield  {author} {\bibinfo {author} {\bibfnamefont {K.F.}\ \bibnamefont
  {Mak}}, \bibinfo {author} {\bibfnamefont {K.}~\bibnamefont {He}}, \bibinfo
  {author} {\bibfnamefont {C.}~\bibnamefont {Lee}}, \bibinfo {author}
  {\bibfnamefont {G.H.}\ \bibnamefont {Lee}}, \bibinfo {author} {\bibfnamefont
  {J.}~\bibnamefont {Hone}}, \bibinfo {author} {\bibfnamefont {T.F.}\
  \bibnamefont {Heinz}}, \ and\ \bibinfo {author} {\bibfnamefont
  {J.}~\bibnamefont {Shan}},\ }\bibfield  {title} {\enquote {\bibinfo {title}
  {Tightly bound trions in monolayer {MoS}$_2$},}\ }\href@noop {} {\bibfield
  {journal} {\bibinfo  {journal} {Nat. Mater.}\ }\textbf {\bibinfo {volume}
  {12}},\ \bibinfo {pages} {207--211} (\bibinfo {year} {2013})}\BibitemShut
  {NoStop}%
\bibitem [{\citenamefont {Ma}\ \emph {et~al.}(2021)\citenamefont {Ma},
  \citenamefont {Nguyen}, \citenamefont {Wang}, \citenamefont {Zeng},
  \citenamefont {Watanabe}, \citenamefont {Taniguchi}, \citenamefont
  {MacDonald}, \citenamefont {Mak},\ and\ \citenamefont
  {Shan}}]{ma2021strongly}%
  \BibitemOpen
  \bibfield  {author} {\bibinfo {author} {\bibfnamefont {L.}~\bibnamefont
  {Ma}}, \bibinfo {author} {\bibfnamefont {P.X.}\ \bibnamefont {Nguyen}},
  \bibinfo {author} {\bibfnamefont {Z.}~\bibnamefont {Wang}}, \bibinfo {author}
  {\bibfnamefont {Y.}~\bibnamefont {Zeng}}, \bibinfo {author} {\bibfnamefont
  {K.}~\bibnamefont {Watanabe}}, \bibinfo {author} {\bibfnamefont
  {T.}~\bibnamefont {Taniguchi}}, \bibinfo {author} {\bibfnamefont {A.H.}\
  \bibnamefont {MacDonald}}, \bibinfo {author} {\bibfnamefont {K.F.}\
  \bibnamefont {Mak}}, \ and\ \bibinfo {author} {\bibfnamefont
  {J.}~\bibnamefont {Shan}},\ }\bibfield  {title} {\enquote {\bibinfo {title}
  {Strongly correlated excitonic insulator in atomic double layers},}\
  }\href@noop {} {\bibfield  {journal} {\bibinfo  {journal} {Nature}\ }\textbf
  {\bibinfo {volume} {598}},\ \bibinfo {pages} {585--589} (\bibinfo {year}
  {2021})}\BibitemShut {NoStop}%
\bibitem [{\citenamefont {Zhang}\ \emph {et~al.}(2025)\citenamefont {Zhang},
  \citenamefont {Nguyen}, \citenamefont {Batra}, \citenamefont {Liu},
  \citenamefont {Watanabe}, \citenamefont {Taniguchi}, \citenamefont
  {Feldman},\ and\ \citenamefont {Li}}]{zhang2025excitons}%
  \BibitemOpen
  \bibfield  {author} {\bibinfo {author} {\bibfnamefont {N.J.}\ \bibnamefont
  {Zhang}}, \bibinfo {author} {\bibfnamefont {R.Q.}\ \bibnamefont {Nguyen}},
  \bibinfo {author} {\bibfnamefont {N.}~\bibnamefont {Batra}}, \bibinfo
  {author} {\bibfnamefont {X.}~\bibnamefont {Liu}}, \bibinfo {author}
  {\bibfnamefont {K.}~\bibnamefont {Watanabe}}, \bibinfo {author}
  {\bibfnamefont {T.}~\bibnamefont {Taniguchi}}, \bibinfo {author}
  {\bibfnamefont {D.E.}\ \bibnamefont {Feldman}}, \ and\ \bibinfo {author}
  {\bibfnamefont {J.}~\bibnamefont {Li}},\ }\bibfield  {title} {\enquote
  {\bibinfo {title} {Excitons in the fractional quantum hall effect},}\
  }\href@noop {} {\bibfield  {journal} {\bibinfo  {journal} {Nature}\ }\textbf
  {\bibinfo {volume} {637}},\ \bibinfo {pages} {327--332} (\bibinfo {year}
  {2025})}\BibitemShut {NoStop}%
\bibitem [{\citenamefont {Madéo}\ \emph {et~al.}(2020)\citenamefont {Madéo},
  \citenamefont {Man}, \citenamefont {Sahoo}, \citenamefont {Campbell},
  \citenamefont {Pareek}, \citenamefont {Wong}, \citenamefont {Al-Mahboob},
  \citenamefont {Chan}, \citenamefont {Karmakar}, \citenamefont {Mariserla}
  \emph {et~al.}}]{madeo2020directly}%
  \BibitemOpen
  \bibfield  {author} {\bibinfo {author} {\bibfnamefont {J.}~\bibnamefont
  {Madéo}}, \bibinfo {author} {\bibfnamefont {M.K.L.}\ \bibnamefont {Man}},
  \bibinfo {author} {\bibfnamefont {C.}~\bibnamefont {Sahoo}}, \bibinfo
  {author} {\bibfnamefont {M.}~\bibnamefont {Campbell}}, \bibinfo {author}
  {\bibfnamefont {V.}~\bibnamefont {Pareek}}, \bibinfo {author} {\bibfnamefont
  {E.L.}\ \bibnamefont {Wong}}, \bibinfo {author} {\bibfnamefont
  {A.}~\bibnamefont {Al-Mahboob}}, \bibinfo {author} {\bibfnamefont {N.S.}\
  \bibnamefont {Chan}}, \bibinfo {author} {\bibfnamefont {A.}~\bibnamefont
  {Karmakar}}, \bibinfo {author} {\bibfnamefont {B.M.K.}\ \bibnamefont
  {Mariserla}},  \emph {et~al.},\ }\bibfield  {title} {\enquote {\bibinfo
  {title} {Directly visualizing the momentum-forbidden dark excitons and their
  dynamics in atomically thin semiconductors},}\ }\href@noop {} {\bibfield
  {journal} {\bibinfo  {journal} {Science}\ }\textbf {\bibinfo {volume}
  {370}},\ \bibinfo {pages} {1199--1204} (\bibinfo {year} {2020})}\BibitemShut
  {NoStop}%
\bibitem [{\citenamefont {Man}\ \emph {et~al.}(2021)\citenamefont {Man},
  \citenamefont {Madéo}, \citenamefont {Sahoo}, \citenamefont {Xie},
  \citenamefont {Campbell}, \citenamefont {Pareek}, \citenamefont {Karmakar},
  \citenamefont {Wong}, \citenamefont {Al-Mahboob}, \citenamefont {Chan} \emph
  {et~al.}}]{man2021experimental}%
  \BibitemOpen
  \bibfield  {author} {\bibinfo {author} {\bibfnamefont {M.K.L.}\ \bibnamefont
  {Man}}, \bibinfo {author} {\bibfnamefont {J.}~\bibnamefont {Madéo}},
  \bibinfo {author} {\bibfnamefont {C.}~\bibnamefont {Sahoo}}, \bibinfo
  {author} {\bibfnamefont {K.}~\bibnamefont {Xie}}, \bibinfo {author}
  {\bibfnamefont {M.}~\bibnamefont {Campbell}}, \bibinfo {author}
  {\bibfnamefont {V.}~\bibnamefont {Pareek}}, \bibinfo {author} {\bibfnamefont
  {A.}~\bibnamefont {Karmakar}}, \bibinfo {author} {\bibfnamefont {E.L.}\
  \bibnamefont {Wong}}, \bibinfo {author} {\bibfnamefont {A.}~\bibnamefont
  {Al-Mahboob}}, \bibinfo {author} {\bibfnamefont {N.S.}\ \bibnamefont {Chan}},
   \emph {et~al.},\ }\bibfield  {title} {\enquote {\bibinfo {title}
  {Experimental measurement of the intrinsic excitonic wave function},}\
  }\href@noop {} {\bibfield  {journal} {\bibinfo  {journal} {Sci. Adv.}\
  }\textbf {\bibinfo {volume} {7}},\ \bibinfo {pages} {eabg0192} (\bibinfo
  {year} {2021})}\BibitemShut {NoStop}%
\bibitem [{\citenamefont {Wallauer}\ \emph {et~al.}(2021)\citenamefont
  {Wallauer}, \citenamefont {Perea-Causin}, \citenamefont {Münster},
  \citenamefont {Zajusch}, \citenamefont {Brem}, \citenamefont {Güdde},
  \citenamefont {Tanimura}, \citenamefont {Lin}, \citenamefont {Huber},
  \citenamefont {Malic} \emph {et~al.}}]{wallauer2021momentum}%
  \BibitemOpen
  \bibfield  {author} {\bibinfo {author} {\bibfnamefont {R.}~\bibnamefont
  {Wallauer}}, \bibinfo {author} {\bibfnamefont {R.}~\bibnamefont
  {Perea-Causin}}, \bibinfo {author} {\bibfnamefont {L.}~\bibnamefont
  {Münster}}, \bibinfo {author} {\bibfnamefont {S.}~\bibnamefont {Zajusch}},
  \bibinfo {author} {\bibfnamefont {S.}~\bibnamefont {Brem}}, \bibinfo {author}
  {\bibfnamefont {J.}~\bibnamefont {Güdde}}, \bibinfo {author} {\bibfnamefont
  {K.}~\bibnamefont {Tanimura}}, \bibinfo {author} {\bibfnamefont {K.Q.}\
  \bibnamefont {Lin}}, \bibinfo {author} {\bibfnamefont {R.}~\bibnamefont
  {Huber}}, \bibinfo {author} {\bibfnamefont {E.}~\bibnamefont {Malic}},  \emph
  {et~al.},\ }\bibfield  {title} {\enquote {\bibinfo {title} {Momentum-resolved
  observation of exciton formation dynamics in monolayer {WS}$_2$},}\
  }\href@noop {} {\bibfield  {journal} {\bibinfo  {journal} {Nano Lett.}\
  }\textbf {\bibinfo {volume} {21}},\ \bibinfo {pages} {5867--5873} (\bibinfo
  {year} {2021})}\BibitemShut {NoStop}%
\bibitem [{\citenamefont {Schmitt}\ \emph {et~al.}(2022)\citenamefont
  {Schmitt}, \citenamefont {Bange}, \citenamefont {Bennecke}, \citenamefont
  {AlMutairi}, \citenamefont {Meneghini}, \citenamefont {Watanabe},
  \citenamefont {Taniguchi}, \citenamefont {Steil}, \citenamefont {Luke},
  \citenamefont {Weitz} \emph {et~al.}}]{schmitt2022formation}%
  \BibitemOpen
  \bibfield  {author} {\bibinfo {author} {\bibfnamefont {D.}~\bibnamefont
  {Schmitt}}, \bibinfo {author} {\bibfnamefont {J.P.}\ \bibnamefont {Bange}},
  \bibinfo {author} {\bibfnamefont {W.}~\bibnamefont {Bennecke}}, \bibinfo
  {author} {\bibfnamefont {A.}~\bibnamefont {AlMutairi}}, \bibinfo {author}
  {\bibfnamefont {G.}~\bibnamefont {Meneghini}}, \bibinfo {author}
  {\bibfnamefont {K.}~\bibnamefont {Watanabe}}, \bibinfo {author}
  {\bibfnamefont {T.}~\bibnamefont {Taniguchi}}, \bibinfo {author}
  {\bibfnamefont {D.}~\bibnamefont {Steil}}, \bibinfo {author} {\bibfnamefont
  {D.R.}\ \bibnamefont {Luke}}, \bibinfo {author} {\bibfnamefont {R.T.}\
  \bibnamefont {Weitz}},  \emph {et~al.},\ }\bibfield  {title} {\enquote
  {\bibinfo {title} {Formation of moiré interlayer excitons in space and
  time},}\ }\href@noop {} {\bibfield  {journal} {\bibinfo  {journal} {Nature}\
  }\textbf {\bibinfo {volume} {608}},\ \bibinfo {pages} {499--503} (\bibinfo
  {year} {2022})}\BibitemShut {NoStop}%
\bibitem [{\citenamefont {Mori}\ \emph {et~al.}(2023)\citenamefont {Mori},
  \citenamefont {Ciocys}, \citenamefont {Takasan}, \citenamefont {Ai},
  \citenamefont {Currier}, \citenamefont {Morimoto}, \citenamefont {Moore},\
  and\ \citenamefont {Lanzara}}]{mori2023spin}%
  \BibitemOpen
  \bibfield  {author} {\bibinfo {author} {\bibfnamefont {R.}~\bibnamefont
  {Mori}}, \bibinfo {author} {\bibfnamefont {S.}~\bibnamefont {Ciocys}},
  \bibinfo {author} {\bibfnamefont {K.}~\bibnamefont {Takasan}}, \bibinfo
  {author} {\bibfnamefont {P.}~\bibnamefont {Ai}}, \bibinfo {author}
  {\bibfnamefont {K.}~\bibnamefont {Currier}}, \bibinfo {author} {\bibfnamefont
  {T.}~\bibnamefont {Morimoto}}, \bibinfo {author} {\bibfnamefont {J.E.}\
  \bibnamefont {Moore}}, \ and\ \bibinfo {author} {\bibfnamefont
  {A.}~\bibnamefont {Lanzara}},\ }\bibfield  {title} {\enquote {\bibinfo
  {title} {Spin-polarized spatially indirect excitons in a topological
  insulator},}\ }\href@noop {} {\bibfield  {journal} {\bibinfo  {journal}
  {Nature}\ }\textbf {\bibinfo {volume} {614}},\ \bibinfo {pages} {249--255}
  (\bibinfo {year} {2023})}\BibitemShut {NoStop}%
\bibitem [{\citenamefont {Zhang}\ \emph {et~al.}(2016)\citenamefont {Zhang},
  \citenamefont {Ugeda}, \citenamefont {Jin}, \citenamefont {Shi},
  \citenamefont {Bradley}, \citenamefont {Martín-Recio}, \citenamefont {Ryu},
  \citenamefont {Kim}, \citenamefont {Tang}, \citenamefont {Kim} \emph
  {et~al.}}]{zhang2016electronic}%
  \BibitemOpen
  \bibfield  {author} {\bibinfo {author} {\bibfnamefont {Y.}~\bibnamefont
  {Zhang}}, \bibinfo {author} {\bibfnamefont {M.M.}\ \bibnamefont {Ugeda}},
  \bibinfo {author} {\bibfnamefont {C.}~\bibnamefont {Jin}}, \bibinfo {author}
  {\bibfnamefont {S.F.}\ \bibnamefont {Shi}}, \bibinfo {author} {\bibfnamefont
  {A.J.}\ \bibnamefont {Bradley}}, \bibinfo {author} {\bibfnamefont
  {A.}~\bibnamefont {Martín-Recio}}, \bibinfo {author} {\bibfnamefont
  {H.}~\bibnamefont {Ryu}}, \bibinfo {author} {\bibfnamefont {J.}~\bibnamefont
  {Kim}}, \bibinfo {author} {\bibfnamefont {S.}~\bibnamefont {Tang}}, \bibinfo
  {author} {\bibfnamefont {Y.}~\bibnamefont {Kim}},  \emph {et~al.},\
  }\bibfield  {title} {\enquote {\bibinfo {title} {Electronic structure,
  surface doping, and optical response in epitaxial {WSe}$_2$ thin films},}\
  }\href@noop {} {\bibfield  {journal} {\bibinfo  {journal} {Nano Lett.}\
  }\textbf {\bibinfo {volume} {16}},\ \bibinfo {pages} {2485--2491} (\bibinfo
  {year} {2016})}\BibitemShut {NoStop}%
\bibitem [{\citenamefont {Wan}\ \emph {et~al.}(2025)\citenamefont {Wan},
  \citenamefont {Zhao}, \citenamefont {Dong}, \citenamefont {Li}, \citenamefont
  {Yang}, \citenamefont {Wang}, \citenamefont {Huang}, \citenamefont {Wen},
  \citenamefont {Li}, \citenamefont {He} \emph {et~al.}}]{wan2025sensitive}%
  \BibitemOpen
  \bibfield  {author} {\bibinfo {author} {\bibfnamefont {Q.}~\bibnamefont
  {Wan}}, \bibinfo {author} {\bibfnamefont {K.}~\bibnamefont {Zhao}}, \bibinfo
  {author} {\bibfnamefont {G.}~\bibnamefont {Dong}}, \bibinfo {author}
  {\bibfnamefont {E.}~\bibnamefont {Li}}, \bibinfo {author} {\bibfnamefont
  {T.}~\bibnamefont {Yang}}, \bibinfo {author} {\bibfnamefont {H.}~\bibnamefont
  {Wang}}, \bibinfo {author} {\bibfnamefont {Y.}~\bibnamefont {Huang}},
  \bibinfo {author} {\bibfnamefont {Y.}~\bibnamefont {Wen}}, \bibinfo {author}
  {\bibfnamefont {Y.}~\bibnamefont {Li}}, \bibinfo {author} {\bibfnamefont
  {J.}~\bibnamefont {He}},  \emph {et~al.},\ }\bibfield  {title} {\enquote
  {\bibinfo {title} {Sensitive infrared surface photovoltage in
  quasi-equilibrium in a layered semiconductor at low-intensity low-temperature
  condition},}\ }\href@noop {} {\bibfield  {journal} {\bibinfo  {journal}
  {arXiv preprint arXiv:2507.08279}\ } (\bibinfo {year} {2025})}\BibitemShut
  {NoStop}%
\bibitem [{\citenamefont {Katoch}\ \emph {et~al.}(2018)\citenamefont {Katoch},
  \citenamefont {Ulstrup}, \citenamefont {Koch}, \citenamefont {Moser},
  \citenamefont {McCreary}, \citenamefont {Singh}, \citenamefont {Xu},
  \citenamefont {Jonker}, \citenamefont {Kawakami}, \citenamefont {Bostwick}
  \emph {et~al.}}]{katoch2018giant}%
  \BibitemOpen
  \bibfield  {author} {\bibinfo {author} {\bibfnamefont {J.}~\bibnamefont
  {Katoch}}, \bibinfo {author} {\bibfnamefont {S.}~\bibnamefont {Ulstrup}},
  \bibinfo {author} {\bibfnamefont {R.J.}\ \bibnamefont {Koch}}, \bibinfo
  {author} {\bibfnamefont {S.}~\bibnamefont {Moser}}, \bibinfo {author}
  {\bibfnamefont {K.M.}\ \bibnamefont {McCreary}}, \bibinfo {author}
  {\bibfnamefont {S.}~\bibnamefont {Singh}}, \bibinfo {author} {\bibfnamefont
  {J.}~\bibnamefont {Xu}}, \bibinfo {author} {\bibfnamefont {B.T.}\
  \bibnamefont {Jonker}}, \bibinfo {author} {\bibfnamefont {R.K.}\ \bibnamefont
  {Kawakami}}, \bibinfo {author} {\bibfnamefont {A.}~\bibnamefont {Bostwick}},
  \emph {et~al.},\ }\bibfield  {title} {\enquote {\bibinfo {title} {Giant
  spin-splitting and gap renormalization driven by trions in single-layer
  {WS}$_2$/h-{BN} heterostructures},}\ }\href@noop {} {\bibfield  {journal}
  {\bibinfo  {journal} {Nat. Phys.}\ }\textbf {\bibinfo {volume} {14}},\
  \bibinfo {pages} {355--359} (\bibinfo {year} {2018})}\BibitemShut {NoStop}%
\bibitem [{\citenamefont {Ma}\ \emph {et~al.}(2022)\citenamefont {Ma},
  \citenamefont {Nie}, \citenamefont {Gui} \emph {et~al.}}]{ma2022}%
  \BibitemOpen
  \bibfield  {author} {\bibinfo {author} {\bibfnamefont {J.}~\bibnamefont
  {Ma}}, \bibinfo {author} {\bibfnamefont {S.}~\bibnamefont {Nie}}, \bibinfo
  {author} {\bibfnamefont {X.}~\bibnamefont {Gui}},  \emph {et~al.},\
  }\bibfield  {title} {\enquote {\bibinfo {title} {Multiple mobile excitons
  manifested as sidebands in quasi-one-dimensional metallic {TaSe}$_3$},}\
  }\href@noop {} {\bibfield  {journal} {\bibinfo  {journal} {Nat. Mater.}\
  }\textbf {\bibinfo {volume} {21}},\ \bibinfo {pages} {423--429} (\bibinfo
  {year} {2022})}\BibitemShut {NoStop}%
\bibitem [{\citenamefont {Rustagi}\ and\ \citenamefont
  {Kemper}(2018)}]{rustagi2018photoemission}%
  \BibitemOpen
  \bibfield  {author} {\bibinfo {author} {\bibfnamefont {A.}~\bibnamefont
  {Rustagi}}\ and\ \bibinfo {author} {\bibfnamefont {A.F.}\ \bibnamefont
  {Kemper}},\ }\bibfield  {title} {\enquote {\bibinfo {title} {Photoemission
  signature of excitons},}\ }\href@noop {} {\bibfield  {journal} {\bibinfo
  {journal} {Phys. Rev. B}\ }\textbf {\bibinfo {volume} {97}},\ \bibinfo
  {pages} {235310} (\bibinfo {year} {2018})}\BibitemShut {NoStop}%
\bibitem [{\citenamefont {Mo}\ \emph {et~al.}(2025)\citenamefont {Mo},
  \citenamefont {Bai}, \citenamefont {Wu}, \citenamefont {Cui}, \citenamefont
  {Mei}, \citenamefont {Wan}, \citenamefont {Li}, \citenamefont {Peng},
  \citenamefont {Zhao}, \citenamefont {Qin} \emph
  {et~al.}}]{mo2025observation}%
  \BibitemOpen
  \bibfield  {author} {\bibinfo {author} {\bibfnamefont {S.}~\bibnamefont
  {Mo}}, \bibinfo {author} {\bibfnamefont {Y.}~\bibnamefont {Bai}}, \bibinfo
  {author} {\bibfnamefont {C.}~\bibnamefont {Wu}}, \bibinfo {author}
  {\bibfnamefont {X.}~\bibnamefont {Cui}}, \bibinfo {author} {\bibfnamefont
  {G.}~\bibnamefont {Mei}}, \bibinfo {author} {\bibfnamefont {Q.}~\bibnamefont
  {Wan}}, \bibinfo {author} {\bibfnamefont {R.}~\bibnamefont {Li}}, \bibinfo
  {author} {\bibfnamefont {C.}~\bibnamefont {Peng}}, \bibinfo {author}
  {\bibfnamefont {K.}~\bibnamefont {Zhao}}, \bibinfo {author} {\bibfnamefont
  {D.}~\bibnamefont {Qin}},  \emph {et~al.},\ }\bibfield  {title} {\enquote
  {\bibinfo {title} {Observation of quasi-steady dark excitons and gap phase in
  a doped semiconductor},}\ }\href@noop {} {\bibfield  {journal} {\bibinfo
  {journal} {arXiv preprint arXiv:2507.08419}\ } (\bibinfo {year}
  {2025})}\BibitemShut {NoStop}%
\bibitem [{SI()}]{SI}%
  \BibitemOpen
  \href@noop {} {}\bibinfo {note} {See Supplemental Material for details on
  other experimental data, data analysis, theoretical calculations and
  simulation of ARPES spectra for exciton, which includes Refs. [12, 13, 17,
  23, 24]}\BibitemShut {NoStop}%
\bibitem [{\citenamefont {Chand}\ \emph {et~al.}(2023)\citenamefont {Chand},
  \citenamefont {Woods}, \citenamefont {Quan}, \citenamefont {Mejia},
  \citenamefont {Taniguchi}, \citenamefont {Watanabe}, \citenamefont {Alù},\
  and\ \citenamefont {Grosso}}]{chand2023interaction}%
  \BibitemOpen
  \bibfield  {author} {\bibinfo {author} {\bibfnamefont {S.B.}\ \bibnamefont
  {Chand}}, \bibinfo {author} {\bibfnamefont {J.M.}\ \bibnamefont {Woods}},
  \bibinfo {author} {\bibfnamefont {J.}~\bibnamefont {Quan}}, \bibinfo {author}
  {\bibfnamefont {E.}~\bibnamefont {Mejia}}, \bibinfo {author} {\bibfnamefont
  {T.}~\bibnamefont {Taniguchi}}, \bibinfo {author} {\bibfnamefont
  {K.}~\bibnamefont {Watanabe}}, \bibinfo {author} {\bibfnamefont
  {A.}~\bibnamefont {Alù}}, \ and\ \bibinfo {author} {\bibfnamefont
  {G.}~\bibnamefont {Grosso}},\ }\bibfield  {title} {\enquote {\bibinfo {title}
  {Interaction-driven transport of dark excitons in 2{D} semiconductors with
  phonon-mediated optical readout},}\ }\href@noop {} {\bibfield  {journal}
  {\bibinfo  {journal} {Nat. Commun.}\ }\textbf {\bibinfo {volume} {14}},\
  \bibinfo {pages} {3712} (\bibinfo {year} {2023})}\BibitemShut {NoStop}%
\bibitem [{\citenamefont {Wu}\ \emph {et~al.}(2019)\citenamefont {Wu},
  \citenamefont {Cheng},\ and\ \citenamefont {Wang}}]{wu2019exciton}%
  \BibitemOpen
  \bibfield  {author} {\bibinfo {author} {\bibfnamefont {S.}~\bibnamefont
  {Wu}}, \bibinfo {author} {\bibfnamefont {L.}~\bibnamefont {Cheng}}, \ and\
  \bibinfo {author} {\bibfnamefont {Q.}~\bibnamefont {Wang}},\ }\bibfield
  {title} {\enquote {\bibinfo {title} {Exciton states and absorption spectra in
  freestanding monolayer transition metal dichalcogenides: A variationally
  optimized diagonalization method},}\ }\href@noop {} {\bibfield  {journal}
  {\bibinfo  {journal} {Phys. Rev. B}\ }\textbf {\bibinfo {volume} {100}},\
  \bibinfo {pages} {115430} (\bibinfo {year} {2019})}\BibitemShut {NoStop}%
\bibitem [{\citenamefont {He}\ \emph {et~al.}(2014)\citenamefont {He},
  \citenamefont {Kumar}, \citenamefont {Zhao}, \citenamefont {Wang},
  \citenamefont {Mak}, \citenamefont {Zhao},\ and\ \citenamefont
  {Shan}}]{he2014tightly}%
  \BibitemOpen
  \bibfield  {author} {\bibinfo {author} {\bibfnamefont {K.}~\bibnamefont
  {He}}, \bibinfo {author} {\bibfnamefont {N.}~\bibnamefont {Kumar}}, \bibinfo
  {author} {\bibfnamefont {L.}~\bibnamefont {Zhao}}, \bibinfo {author}
  {\bibfnamefont {Z.}~\bibnamefont {Wang}}, \bibinfo {author} {\bibfnamefont
  {K.F.}\ \bibnamefont {Mak}}, \bibinfo {author} {\bibfnamefont
  {H.}~\bibnamefont {Zhao}}, \ and\ \bibinfo {author} {\bibfnamefont
  {J.}~\bibnamefont {Shan}},\ }\bibfield  {title} {\enquote {\bibinfo {title}
  {Tightly bound excitons in monolayer {WSe}$_2$},}\ }\href@noop {} {\bibfield
  {journal} {\bibinfo  {journal} {Phys. Rev. Lett.}\ }\textbf {\bibinfo
  {volume} {113}},\ \bibinfo {pages} {026803} (\bibinfo {year}
  {2014})}\BibitemShut {NoStop}%
\bibitem [{\citenamefont {Riley}\ \emph {et~al.}(2028)\citenamefont {Riley},
  \citenamefont {Caruso}, \citenamefont {Verdi}, \citenamefont {Duffy},
  \citenamefont {Watson}, \citenamefont {Bawden}, \citenamefont {Volckaert},
  \citenamefont {Laan}, \citenamefont {Hesjedal}, \citenamefont {Hoesch} \emph
  {et~al.}}]{riley2018crossover}%
  \BibitemOpen
  \bibfield  {author} {\bibinfo {author} {\bibfnamefont {J.M.}\ \bibnamefont
  {Riley}}, \bibinfo {author} {\bibfnamefont {F.}~\bibnamefont {Caruso}},
  \bibinfo {author} {\bibfnamefont {C.}~\bibnamefont {Verdi}}, \bibinfo
  {author} {\bibfnamefont {L.B.}\ \bibnamefont {Duffy}}, \bibinfo {author}
  {\bibfnamefont {M.D.}\ \bibnamefont {Watson}}, \bibinfo {author}
  {\bibfnamefont {L.}~\bibnamefont {Bawden}}, \bibinfo {author} {\bibfnamefont
  {K.}~\bibnamefont {Volckaert}}, \bibinfo {author} {\bibfnamefont
  {G.~Van~Der}\ \bibnamefont {Laan}}, \bibinfo {author} {\bibfnamefont
  {T.}~\bibnamefont {Hesjedal}}, \bibinfo {author} {\bibfnamefont
  {M.}~\bibnamefont {Hoesch}},  \emph {et~al.},\ }\bibfield  {title} {\enquote
  {\bibinfo {title} {Crossover from lattice to plasmonic polarons of a
  spin-polarised electron gas in ferromagnetic {EuO}},}\ }\href@noop {}
  {\bibfield  {journal} {\bibinfo  {journal} {Nat. Commun.}\ }\textbf {\bibinfo
  {volume} {9}},\ \bibinfo {pages} {2305} (\bibinfo {year} {2028})}\BibitemShut
  {NoStop}%
\bibitem [{\citenamefont {Ulstrup}\ \emph {et~al.}(2024)\citenamefont
  {Ulstrup}, \citenamefont {in't Veld}, \citenamefont {Miwa}, \citenamefont
  {Jones}, \citenamefont {McCreary}, \citenamefont {Robinson}, \citenamefont
  {Jonker}, \citenamefont {Singh}, \citenamefont {Koch}, \citenamefont
  {Rotenberg} \emph {et~al.}}]{ulstrup2024observation}%
  \BibitemOpen
  \bibfield  {author} {\bibinfo {author} {\bibfnamefont {S.}~\bibnamefont
  {Ulstrup}}, \bibinfo {author} {\bibfnamefont {Y.}~\bibnamefont {in't Veld}},
  \bibinfo {author} {\bibfnamefont {J.A.}\ \bibnamefont {Miwa}}, \bibinfo
  {author} {\bibfnamefont {A.J.H.}\ \bibnamefont {Jones}}, \bibinfo {author}
  {\bibfnamefont {K.M.}\ \bibnamefont {McCreary}}, \bibinfo {author}
  {\bibfnamefont {J.T.}\ \bibnamefont {Robinson}}, \bibinfo {author}
  {\bibfnamefont {B.T.}\ \bibnamefont {Jonker}}, \bibinfo {author}
  {\bibfnamefont {S.}~\bibnamefont {Singh}}, \bibinfo {author} {\bibfnamefont
  {R.J.}\ \bibnamefont {Koch}}, \bibinfo {author} {\bibfnamefont
  {E.}~\bibnamefont {Rotenberg}},  \emph {et~al.},\ }\bibfield  {title}
  {\enquote {\bibinfo {title} {Observation of interlayer plasmon polaron in
  graphene/{WS}$_2$ heterostructures},}\ }\href@noop {} {\bibfield  {journal}
  {\bibinfo  {journal} {Nat. Commun.}\ }\textbf {\bibinfo {volume} {15}},\
  \bibinfo {pages} {3845} (\bibinfo {year} {2024})}\BibitemShut {NoStop}%
\bibitem [{\citenamefont {Zhang}\ \emph {et~al.}(2014)\citenamefont {Zhang},
  \citenamefont {Richard} \emph {et~al.}}]{zhang2014observation}%
  \BibitemOpen
  \bibfield  {author} {\bibinfo {author} {\bibfnamefont {P.}~\bibnamefont
  {Zhang}}, \bibinfo {author} {\bibfnamefont {P.}~\bibnamefont {Richard}},
  \emph {et~al.},\ }\bibfield  {title} {\enquote {\bibinfo {title} {Observation
  of momentum-confined in-gap impurity state in
  {Ba}$_{0.6}${K}$_{0.4}${Fe}$_2${As}$_2$: Evidence for antiphase
  s$\pm$pairing},}\ }\href@noop {} {\bibfield  {journal} {\bibinfo  {journal}
  {Phys. Rev. X}\ }\textbf {\bibinfo {volume} {4}},\ \bibinfo {pages} {031001}
  (\bibinfo {year} {2014})}\BibitemShut {NoStop}%
\bibitem [{\citenamefont {Strocov}\ \emph {et~al.}(2018)\citenamefont
  {Strocov}, \citenamefont {Cancellieri},\ and\ \citenamefont
  {Mishchenko}}]{Strocov2018}%
  \BibitemOpen
  \bibfield  {author} {\bibinfo {author} {\bibfnamefont {V.N.}\ \bibnamefont
  {Strocov}}, \bibinfo {author} {\bibfnamefont {C.}~\bibnamefont
  {Cancellieri}}, \ and\ \bibinfo {author} {\bibfnamefont {A.S.}\ \bibnamefont
  {Mishchenko}},\ }\enquote {\bibinfo {title} {Electrons and polarons at oxide
  interfaces explored by soft-x-ray arpes},}\ in\ \href@noop {} {\emph
  {\bibinfo {booktitle} {Spectroscopy of Complex Oxide Interfaces:
  Photoemission and Related Spectroscopies}}},\ \bibinfo {editor} {edited by\
  \bibinfo {editor} {\bibfnamefont {C.}~\bibnamefont {Cancellieri}}\ and\
  \bibinfo {editor} {\bibfnamefont {V.N.}\ \bibnamefont {Strocov}}}\ (\bibinfo
  {publisher} {Springer International Publishing},\ \bibinfo {address} {Cham},\
  \bibinfo {year} {2018})\ pp.\ \bibinfo {pages} {107--151}\BibitemShut
  {NoStop}%
\bibitem [{\citenamefont {Cercellier}\ \emph {et~al.}(2007)\citenamefont
  {Cercellier}, \citenamefont {Monney}, \citenamefont {Clerc}, \citenamefont
  {Battaglia}, \citenamefont {Despont}, \citenamefont {Garnier}, \citenamefont
  {Beck}, \citenamefont {Aebi}, \citenamefont {Patthey}, \citenamefont {Berger}
  \emph {et~al.}}]{cercellier2007evidence}%
  \BibitemOpen
  \bibfield  {author} {\bibinfo {author} {\bibfnamefont {H.}~\bibnamefont
  {Cercellier}}, \bibinfo {author} {\bibfnamefont {C.}~\bibnamefont {Monney}},
  \bibinfo {author} {\bibfnamefont {F.}~\bibnamefont {Clerc}}, \bibinfo
  {author} {\bibfnamefont {C.}~\bibnamefont {Battaglia}}, \bibinfo {author}
  {\bibfnamefont {L.}~\bibnamefont {Despont}}, \bibinfo {author} {\bibfnamefont
  {M.G.}\ \bibnamefont {Garnier}}, \bibinfo {author} {\bibfnamefont
  {H.}~\bibnamefont {Beck}}, \bibinfo {author} {\bibfnamefont {P.}~\bibnamefont
  {Aebi}}, \bibinfo {author} {\bibfnamefont {L.}~\bibnamefont {Patthey}},
  \bibinfo {author} {\bibfnamefont {H.}~\bibnamefont {Berger}},  \emph
  {et~al.},\ }\bibfield  {title} {\enquote {\bibinfo {title} {Evidence for an
  excitonic insulator phase in 1{T}-{TiSe}$_2$},}\ }\href@noop {} {\bibfield
  {journal} {\bibinfo  {journal} {Phys. Rev. Lett.}\ }\textbf {\bibinfo
  {volume} {99}},\ \bibinfo {pages} {146403} (\bibinfo {year}
  {2007})}\BibitemShut {NoStop}%
\bibitem [{\citenamefont {Gao}\ \emph {et~al.}(2024)\citenamefont {Gao},
  \citenamefont {Chan}, \citenamefont {Jiao}, \citenamefont {Chen},
  \citenamefont {Yin}, \citenamefont {Tangprapha}, \citenamefont {Yang},
  \citenamefont {Li}, \citenamefont {Liu}, \citenamefont {Shen} \emph
  {et~al.}}]{gao2024observation}%
  \BibitemOpen
  \bibfield  {author} {\bibinfo {author} {\bibfnamefont {Q.}~\bibnamefont
  {Gao}}, \bibinfo {author} {\bibfnamefont {Y.}~\bibnamefont {Chan}}, \bibinfo
  {author} {\bibfnamefont {P.}~\bibnamefont {Jiao}}, \bibinfo {author}
  {\bibfnamefont {H.}~\bibnamefont {Chen}}, \bibinfo {author} {\bibfnamefont
  {S.}~\bibnamefont {Yin}}, \bibinfo {author} {\bibfnamefont {K.}~\bibnamefont
  {Tangprapha}}, \bibinfo {author} {\bibfnamefont {Y.}~\bibnamefont {Yang}},
  \bibinfo {author} {\bibfnamefont {X.}~\bibnamefont {Li}}, \bibinfo {author}
  {\bibfnamefont {Z.}~\bibnamefont {Liu}}, \bibinfo {author} {\bibfnamefont
  {D.}~\bibnamefont {Shen}},  \emph {et~al.},\ }\bibfield  {title} {\enquote
  {\bibinfo {title} {Observation of possible excitonic charge density waves and
  metal--insulator transitions in atomically thin semimetals},}\ }\href@noop {}
  {\bibfield  {journal} {\bibinfo  {journal} {Nat. Phys.}\ }\textbf {\bibinfo
  {volume} {20}},\ \bibinfo {pages} {597--602} (\bibinfo {year}
  {2024})}\BibitemShut {NoStop}%
\bibitem [{\citenamefont {Monney}\ \emph {et~al.}(2009)\citenamefont {Monney},
  \citenamefont {Cercellier}, \citenamefont {Clerc}, \citenamefont {Battaglia},
  \citenamefont {Schwier}, \citenamefont {Didiot}, \citenamefont {Garnier},
  \citenamefont {Beck}, \citenamefont {Aebi}, \citenamefont {Berger} \emph
  {et~al.}}]{monney2009spontaneous}%
  \BibitemOpen
  \bibfield  {author} {\bibinfo {author} {\bibfnamefont {C.}~\bibnamefont
  {Monney}}, \bibinfo {author} {\bibfnamefont {H.}~\bibnamefont {Cercellier}},
  \bibinfo {author} {\bibfnamefont {F.}~\bibnamefont {Clerc}}, \bibinfo
  {author} {\bibfnamefont {C.}~\bibnamefont {Battaglia}}, \bibinfo {author}
  {\bibfnamefont {E.F.}\ \bibnamefont {Schwier}}, \bibinfo {author}
  {\bibfnamefont {C.}~\bibnamefont {Didiot}}, \bibinfo {author} {\bibfnamefont
  {M.G.}\ \bibnamefont {Garnier}}, \bibinfo {author} {\bibfnamefont
  {H.}~\bibnamefont {Beck}}, \bibinfo {author} {\bibfnamefont {P.}~\bibnamefont
  {Aebi}}, \bibinfo {author} {\bibfnamefont {H.}~\bibnamefont {Berger}},  \emph
  {et~al.},\ }\bibfield  {title} {\enquote {\bibinfo {title} {Spontaneous
  exciton condensation in 1{T}-{TiSe}$_2$: Bcs-like approach},}\ }\href@noop {}
  {\bibfield  {journal} {\bibinfo  {journal} {Phys. Rev. B}\ }\textbf {\bibinfo
  {volume} {79}},\ \bibinfo {pages} {045116} (\bibinfo {year}
  {2009})}\BibitemShut {NoStop}%
\bibitem [{\citenamefont {Mishra}\ \emph {et~al.}(2018)\citenamefont {Mishra},
  \citenamefont {Bose}, \citenamefont {Dhar},\ and\ \citenamefont
  {Bhattacharya}}]{mishra2018exciton}%
  \BibitemOpen
  \bibfield  {author} {\bibinfo {author} {\bibfnamefont {H.}~\bibnamefont
  {Mishra}}, \bibinfo {author} {\bibfnamefont {A.}~\bibnamefont {Bose}},
  \bibinfo {author} {\bibfnamefont {A.}~\bibnamefont {Dhar}}, \ and\ \bibinfo
  {author} {\bibfnamefont {S.}~\bibnamefont {Bhattacharya}},\ }\bibfield
  {title} {\enquote {\bibinfo {title} {Exciton-phonon coupling and band-gap
  renormalization in monolayer {WSe}$_2$},}\ }\href@noop {} {\bibfield
  {journal} {\bibinfo  {journal} {Phys. Rev. B}\ }\textbf {\bibinfo {volume}
  {98}},\ \bibinfo {pages} {045143} (\bibinfo {year} {2018})}\BibitemShut
  {NoStop}%
\end{thebibliography}%

\end{document}